\documentclass{article}
\usepackage{amsmath,amssymb,amsfonts,graphicx,bm}
\usepackage{pgfplots}

\renewcommand{\r}{\mathbf{r}}
\newcommand{\dis}{\displaystyle}

\newcommand{\x}{\mathbf{x}}
\newcommand{\n}{\mathbf{n}}
\newcommand{\y}{\mathbf{y}}

\newcommand{\calU}{\mathcal{U}}

\newcommand{\R}{\mathbb{R}}

\begin{document}

\title{Asymptotic analysis of conversion-limited phase separation}

\author{\em Paul C. Bressloff,\\
Department of Mathematics, Imperial College London, \\
London SW7 2AZ, UK.}
\maketitle

\begin{abstract}
Liquid-liquid phase separation plays a major role in the formation and maintenance of various membrane-less subcellular structures in the cytoplasm and nucleus of cells. Biological condensates contain enhanced concentrations of proteins and RNA, many of which can be continually exchanged with the surrounding medium. Coarsening is an important step in the kinetics of phase separation, whereby an emulsion of polydisperse condensates transitions to a single condensate in thermodynamic equilibrium with a surrounding dilute phase. A key feature of biological phase separation is the co-existence of multiple condensates over significant time scales, which is consistent with experimental observations showing a slowing of coarsening rates. It has recently been proposed that one rate limiting step could be the slow interfacial conversion of a molecular constituent between the dilute and dense phases. In this paper we analyze conversion-limited phase separation within the framework of diffusion in singularly perturbed domains, which exploits the fact that biological condensates tend to be much smaller than the size of a cell. Using matched asymptotic analysis, we solve the quasi-static diffusion equation for the concentration in the dilute phase, and then derive kinetic equations for the slow growth/shrinkage of the condensates. This provides a systematic way of obtaining corrections to mean-field theory that take into account the geometry of the cell and the locations of all the condensates.

 \end{abstract}

 \section{Introduction}

There are many phenomena in cell biology that can be characterized in terms of diffusion in a singularly perturbed domain, that is, a bounded region $\Omega \subset \R^d$ that contains a set of small obstacles or traps $\calU_k \subset \Omega$,
$k=1,\ldots,N$, with $|\calU_k|\sim \epsilon^d|\Omega|$ and $0<\epsilon \ll 1$ \cite{PCBbook}. Examples at the single cell level include receptor trafficking in the dendrites of neurons \cite{Bressloff08,Schumm22}, and T cell searching behavior in lymph nodes \cite{Coombs15}. The traps correspond to postsynaptic target domains and antigen presenting cells, respectively. An important example at the multi-cellular level is bacterial quorum sensing, in which autoinducer signaling molecules diffuse extracellularly and the traps are individual bacterial cells \cite{Muller13,Gou16}. Typical quantities of interest at the level of bulk diffusion include the steady-state solution (if it exists) and the long-time rate of relaxation to steady state. On the other hand, at the single-particle level, the diffusion equation represents the evolution of a probability density rather than a macroscopic particle concentration. One is now typically interested in the splitting probabilities and conditional first passage times (FPTs) to be absorbed by one of the traps. 

\begin{figure}[t!]
\centering
\includegraphics[width=12cm]{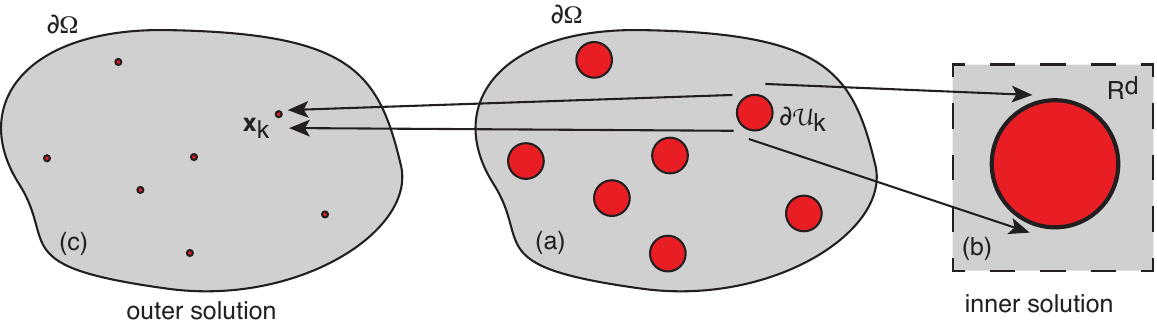} 
\caption{(a) Example of a singularly perturbed domain $\Omega \subset \R^d$ containing $N$ small traps $\calU_k$, $k=1,\ldots,N$, with reactive boundaries $\partial \calU_k$. The small trap limit is $\epsilon \rightarrow 0$, where $|\calU_k| \sim\epsilon^d |\Omega|$. (b) The inner solution of the corresponding diffusion equation is constructed in terms of stretched coordinates $\y=\epsilon^{-1}(\x-{\x}_k)$, where ${\x}_k$ is the center of the $k$-th trap. (c) This is matched with an outer solution by shrinking each trap to a single point and expressing the outer solution in terms of the Green's function for diffusion in $\Omega$. [The diagram is not to scale]. }
\label{fig1}
\end{figure}

In all of the above cases, the resulting boundary value problem (BVP) can be solved using a combination of matched asymptotic analysis and Green's function methods \cite{Ward93,Ward93a,Straube07,Coombs09,Cheviakov11,Chevalier11,Ward15,Lindsay16,Lindsay17,Grebenkov20,Bressloff21a,Bressloff21b,Bressloff22b}.  A schematic diagram of the analytical procedure is shown in Fig. \ref{fig1}. The first step is to obtain an inner solution around each trap by introducing local stretched coordinates, and solving a simplified diffusion equation in which the exterior boundary $\partial \Omega$ and other traps are ignored. This is then matched with an outer solution that satisfies the exterior boundary condition together with a set of singularity conditions at the trap centers, which are determined by the matching procedure. Matching is achieved by writing the outer solution in terms of the corresponding Green's function in the absence of traps. It follows that the details of the asymptotic analysis depend on the spatial dimension $d$ of the bounded domain $\Omega\subset \R^d$, since the singular nature of the Green's function $G(\x,\x')$ for $\x,\x'\in \Omega$ depends on $d$. In particular, the two-dimensional (2D) Green's function varies as $-\log r$ whereas the three-dimensional Green's function varies as $1/r$ as $r\rightarrow 0$ with $r=|\x-\x'|$. Assuming that the exterior boundary $\partial \Omega$ is totally reflecting, the outer solution is approximately constant away from the traps but picks up spatially varying terms in a boundary layer around each trap in order to satisfy the appropriate local boundary condition. These additional terms depend on the geometry of the domain $\Omega$, and the locations and shapes of all the traps.

Recently, we applied the theory of diffusion in singularly perturbed domains to a non-equilibrium statistical physics problem involving liquid-liquid phase separation \cite{Bressloff20a,Bressloff20b}. More specifically, we considered the coexistence of multiple subcellular structures known colloquially as biological condensates. There is now considerable experimental evidence that many different types of biological condensate are found intracellularly in the cytoplasm and cell nucleus (see the reviews \cite{Hyman14,Brangwynne15,Banani17,Berry18,Falahati19,Mittag22} and references therein). These membrane-less organelles are viscous, liquid-like structures containing enhanced concentrations of various proteins and RNA, many of which can be continually exchanged with the surrounding medium. Another example where liquid-liquid phase separation appears to play an important role is in the formation and maintenance of neuronal synapses \cite{Zeng16,Bai21,Hosokawa21}. In this case the biological condensate consists of an enhanced concentration of scaffolding proteins within the cell membrane that can trap diffusing neurotransmitter receptors.

A major stage in the evolution of liquid-liquid phase separation is {\em coarsening}. This is the process whereby an emulsion of polydisperse condensates transitions to a single condensate in thermodynamic equilibrium with a surrounding dilute phase. Late-stage coarsening typically involves two distinct mechanisms \cite{Weber19}: (i) the diffusive transfer of material from small to large droplets via Lisfshitz-Slyozov-Wagner (LSW) Ostwald ripening  \cite{Lif61,Wagner61}; (ii) the coalescence of diffusing droplets through collisions. A key feature of biological liquid-liquid phase separation is the co-existence of multiple condensates over significant time scales, which is consistent with experimental observations showing a slowing of coarsening rates. Various hypotheses have been given to account for this slowing. One example is actively driven chemical reactions that maintain the out-of-equilibrium switching of proteins between soluble and phase separating forms \cite{Zwicker14,Zwicker15,Wurtz18,Lee19}. Another example is the mechanical suppression of coarsening mediated by intracellular visco-elastic networks such as the cytoskeleton \cite{Feric13,Style18,Ros20}. 

Recently, a third possible mechanism has been suggested by an experimental study of the coarsening dynamics of P granules in germ cells of {\em C. elegans} \cite{Folk21}. P granules, which play an important role in asymmetric cell division, are RNA/protein-rich bodies located in a neighborhood of the nucleus \cite{Seydoux18}, and were the first example of biological condensates to be studied quantitatively \cite{Brangwynne09}. In Ref. \cite{Folk21} it was argued that the observed suppression of coarsening is due to the slow conversion of a molecular constituent between the dilute and dense phases, which can be regulated by various protein assemblies that associate with an interface. This motivated the construction of a mean field model of so-called conversion-limited phase separation \cite{Folk21,Lee21}, in which the rate of coarsening is proportional to an interfacial conversion rate $\kappa_0$ rather than the factor $D/R$ of Ostwald ripening, where $D$ is the diffusivity and $R$ is droplet radius. The model supported a significantly slower coarsening rate, consistent with multiple condensates persisting over relevant biological time scales. 

In this paper we analyze the coexistence of multiple condensates under conversion-limited phase separation within the framework of diffusion in singularly perturbed domains. This complements our previous work, which focused on the active suppression of Ostwald ripening due to driven chemical reactions \cite{Bressloff20a,Bressloff20b}. (We neglect coarsening due to coalescence of condensates, since the latter is rarely observed {\em in vivo}.) For the given application, $\Omega$ can be interpreted as the cell interior, say, whereas the local traps represent the condensates. There are several reasons for developing an asymptotic analysis of such a system. First, the small trap assumption is consistent with the observed size of condensates relative to the size of a cell. Second, the constant value of the outer solution away from the traps can be identified with the mean field $\phi_{\infty}$ that determines the supersaturation. The advantage of the asymptotic method is that it provides a systematic way of deriving mean-field theory as well as higher-order corrections that take into account the geometry of the domain $\Omega$ and the locations and shapes of all the condensates (finite-size effects). This leads to a third motivation for our approach, namely, mean field theory breaks down in the case of circular droplets in 2D systems, since the concentration around each condensate varies logarithmically with respect to spatial separation \cite{Kavanagh14,Bressloff20a}.

The structure of the paper is as follow. In section 2, we briefly review the mean-field theory of Ostwald ripening in 3D and then describe the recent model of 3D conversion-limited phase separation introduced in Refs. \cite{Folk21,Lee21}. The asymptotic analysis of the latter is developed in section 3. First, we use a separation of time scales to determine the quasi-static concentration in the dilute phase for fixed droplet radii $R_j$, $j=1,\ldots,N$. We construct an asymptotic expansion of the concentration around each condensate (inner solution) in powers of $\epsilon$ by matching with an outer solution expressed in terms of the 3D Neumann Green's function in $\Omega$. The inner solution is then used to calculate the flux at the interface of each droplet, from which the kinetic equations for droplet growth and shrinkage are derived. We thus establish how coarsening depends on the conversion rate $\kappa_0$ for adsorption/desorption at the droplet interfaces and show that to leading order the supersaturation $\Delta(t)\propto \sum_{j=1}^NR_j(t)/(\sum_{j=1}^NR_j^2(t))$. This contrasts with the classical result $\Delta(t)\propto N/(\sum_{j=1}^NR_j(t))$ for Ostwald ripening in 3D.

In section 4 we carry out the corresponding asymptotic analysis of 2D conversion-limited phase separation. In our previous work on Ostwald ripening \cite{Bressloff20a,Bressloff20b}, see also \cite{Kavanagh14}, we showed that there are major differences between 2D and 3D. Most notably, the 2D supersaturation is proportional to the inverse of the harmonic mean $R_{\rm harm}=N^{-1}\sum_{j=1}^N 1/R_j$ rather than the arithmetic mean. These differences are also reflected in the details of the asymptotic analysis. In the case of 2D Ostwald ripening, the relevant small parameter is $\nu=-1/\ln \epsilon$ rather than $\epsilon$, and matching of the inner and outer solutions involves a non-perturbative summation over all logarithmic singularities.  
On the other hand, the asymptotic analysis of 2D conversion-limited phase separation is based on an asymptotic expansion with respect to $\epsilon$. However, in contrast to the 3D case, higher-order corrections to 2D mean field theory also depend non-perturbatively on the additional small parameter $\delta=-\epsilon \ln \epsilon$. Moreover, the supersaturation has the same form as 3D classical Ostwald ripening. Finally, in section 5 we investigate 3D conversion-limited phase separation in the presence of a spatial concentration gradient for some regulatory protein. In particular, we show how dissolution of condensates can occur at one end of the gradient and assembly of condensates at the other end. This is consistent with P granule segregation observed during zygote polarization \cite{Folk21}.

\section{Ostwald ripening vs. conversion-limited phase separation}

\subsection{Ostwald ripening} 
Consider a set of $N$ condensates $\calU_k$, $k=1,\ldots,N$, within a 3D domain $\Omega\subset \R^3$ that are well separated from each other and whose total volume fraction is relatively small. A no-flux boundary condition on $\partial \Omega$ ensures mass conservation. Each droplet is taken to be a sphere of radius $R_i$ centered about $\x_i \in \Omega$, that is, $\calU_i=\{\x \in \Omega, |\x-\x_i| \leq R_i\}$. We also assume that the coarsening dynamics is much slower than the equilibration of the concentration profile in the dilute phase. Under this quasi-static approximation, the solute concentration $\phi$ exterior to the droplets satisfies a steady-state diffusion equation of the form
\begin{subequations}
\label{Mullins}
\begin{equation}
\nabla^2 \phi=0,\quad \r \in \Omega\backslash \cup_{i=1}^N\calU_{i},\quad \nabla \phi\cdot \n =0 \mbox{ on } \partial \Omega,
\end{equation}
with $\n$ denoting the outward unit normal on $\partial \Omega$, and
\begin{align}
\label{GT}
\phi=\phi_a\left (1+\frac{\ell_c}{R_i}\right )\equiv {\phi}_a(R_i) \mbox{ on }\partial \calU_{i},
\end{align}
\end{subequations}
where $\phi_a$ and $\phi_b$ are the equilibrium concentrations of the dilute and dense phases, respectively, and $\ell_c$ is the capillary length. The boundary condition (\ref{GT}) is known as the Gibbs-Thomson law, and arises from surface tension on the droplet interface. An important consequence of the quasi-static approximation is that the total volume of condensates is conserved. This follows from integrating equation (\ref{Mullins}a) with respect to $\x\in \Omega\backslash \cup_{i=1}^N\calU_{i}$ and using the divergence theorem, which shows that
\begin{equation}
\sum_{j=1}^N \int_{\calU_j}\nabla \phi(\x)\cdot \n d\x=0.
\end{equation}
In other words, the sum of the fluxes into the $N$ condensates is zero so that there is no net change in the total condensate volume.

\begin{figure}[t!]
  \centering
  \includegraphics[width=8cm]{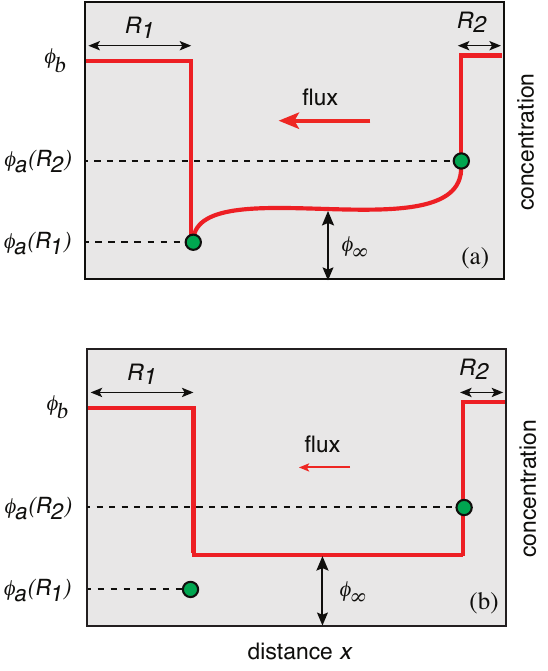}
  \caption{Two distinct coarsening scenarios for a pair of droplets of radii $R_1$ and $R_2$, respectively. (a) Ostwald ripening (diffusion-limited):  the solute concentration ${\phi}_a(R_1)$ around the larger droplet is lower than the concentration $ {\phi}_a(R_2)$ around the smaller droplet, resulting in a net diffusive flux from the small droplet to the large droplet. (b) Conversion-limited coarsening: the rate limiting step is the exchange of solute molecules between the dilute and dense phases across the droplet interfaces. Under a mean-field approximation, the concentration in the dilute phase is assumed to be spatially uniform.}
  \label{fig2}
\end{figure}

In the case of classical coarsening via Ostwald ripening, the difference in surface concentrations $\phi_a(R_i)$ for droplets of different sizes results in a net diffusive flux from small to large droplets. This is illustrated in Fig. \ref{fig2}(a) for two droplets $\calU_1$ and $\calU_2$ with $R_2 < R_1$. Under the mean field approximation, the effects of the boundary $\partial \Omega$ and interactions between droplets are ignored by introducing
a constant mean field $\phi_{\infty}$ with $\phi(\x)\approx \phi_{\infty}$ for all $\x \in \Omega $ such that $|\x-\x_i|\gg R_i$, $i=1,\ldots,N$. The quantity $\Delta=\phi_{\infty}-\phi_a$ is known as the supersaturation, and is determined self-consistently from mass conservation. 
Under the above assumptions, we can focus on a single droplet of radius $R$, whose concentration satisfies the spherically symmetric diffusion equation
 \begin{equation}
0=\frac{D}{r^2}\frac{\partial}{\partial r}\left [r^2\frac{\partial \phi}{\partial r}\right ],\quad r>R,
\end{equation}
together with the boundary conditions
\begin{equation}
\phi(R)= {\phi}_a(R),\quad \phi(r)\rightarrow \phi_{\infty} \mbox{ as } r \rightarrow \infty.
\end{equation}
The solution is given by
\begin{equation}
\label{sol3}
\phi(r)=\phi_{\infty}-\frac{R}{r}\left (\Delta -\frac{\phi_a\ell_c}{R}\right ),
\end{equation}
which implies that the diffusive flux entering the droplet at its interface is
\begin{equation}
J_{R}=D\phi'(R)=\frac{D}{R}\left (\Delta -\frac{\phi_a\ell_c}{R}\right ).
\label{JR}
\end{equation}
We can identify $R_c= \phi_a \ell_c/\Delta$ as a critical radius, with $J_R>0$ ($J_R <0$) when $R>R_c$ ($R<R_c$), so that the droplet grows (shrinks) on longer time scales. In Ref. \cite{Bressloff21a} we showed that equation (\ref{sol3}) can be interpreted as the leading order contribution to the  inner solution in a neighborhood of the droplet, which matches the leading order contribution $\phi_{\infty}$ to the outer solution. However, we were also able to calculate higher-order corrections to the solutions that depended on the geometry of the full system.

Using a separation of time scales, one can now write down dynamical equations for the evolution of the droplet radii. This is driven by the transfer of solute molecules between the dilute and dense phases as determined by the flux. One finds that
\begin{equation}
\frac{dR_i}{dt} =\frac{\Gamma}{R_i}\left (\frac{1}{R_c}-\frac{1}{R_i}\right ),\quad i=1,\ldots,N ,
\label{radOR}
\end{equation}
where
\begin{equation}
\Gamma = \frac{D\phi_a\ell_c}{ (\phi_b-{\phi}_a(R_i))}\approx \frac{D\phi_a\ell_c}{ \phi_b}.
\end{equation}
Multiplying both sides of equation (\ref{radOR}) by $R_i^2$, summing over $i$, and imposing conservation of the total droplet volume $V_{\rm drop} =4\pi \sum_iR_i^3(t)/3 $ gives
\begin{equation}
R_c(t)=\frac{1}{N}\sum_{i=1}^NR_i(t).
\label{Rc0}
\end{equation}
It follows that
\begin{equation}
\label{phin}
\Delta(t)\equiv \phi_{\infty}(t)-\phi_a=\frac{\ell_c\phi_a N}{\sum_{i=1}^NR_i(t)} .
\end{equation}
Equation (\ref{phin}) implies that $\phi_{\infty}(t) $ decreases as the mean radius increases. Since the critical radius $R_c(t)$ increases as the saturation $\Delta(t)=\phi_{\infty}(t)-\phi_a$ decreases, it follows that only a single droplet remains in the limit $t\rightarrow \infty$. 

\subsection{Conversion-limited phase separation} 

The rate limiting step of conversion-limited phase separation is the exchange of solute molecules between the dilute and dense phases across the droplet interfaces \cite{Folk21,Lee21}, see Fig. \ref{fig2}(b). The flux across an interface of radius $R$ is no longer diffusion-limited as in equation (\ref{JR}). Instead, it is taken to have the form
\begin{equation}
J_R=\kappa_0\left (\overline{\phi} -\phi_a-\frac{\phi_a \ell_c}{R}\right ),
\label{JRCL}
\end{equation}
where $\kappa_0$ is a constant (with dimensions of speed) that specifies the effective rates of adsorption and desorption at the interface. In addition, the concentration $\overline{\phi}$ in the dilute phase is assumed to be spatially uniform. 
One physical mechanism for slow absorption and desorption at a droplet interface is the inherent slow internal dynamics within a condensate. The latter could  be due to multi-valent interactions of various protein domains, resulting in a rugged energy landscape that drastically slows down molecular  reconfigurations. (Indeed, it has been shown that P granules act like high viscosity ``aging Maxwell fluids''  that display glass-like properties \cite{Jawerth20}.) It follows that on longer time scales the radii evolve according to the equations
\begin{equation}
\frac{dR_i}{dt} =\kappa\left (\frac{1}{R_c}-\frac{1}{R_i}\right ),\quad i=1,\ldots,N ,
\label{radCL}
\end{equation}
where 
\begin{equation}
\kappa=\frac{\kappa_0\ell_c\phi_a}{\phi_b},\quad  R_c= \frac{\phi_a \ell_c}{\overline{\phi}-\phi_a}.
\end{equation}
Comparison of the kinetic equations (\ref{radCL}) and (\ref{radOR}) establishes one major difference between conversion-limited phase separation and Ostwald ripening: the speed of droplet growth/shrinkage is proportional to $\kappa_0$ in the former case and $D/R_i(t)$ in the latter case. This has two important consequences. First, conversion-limited-coarsening can be much slower than classical Ostwald ripening. Second, the speed of coarsening $\kappa_0$ can be regulated by modifying the polymeric structure of a condensate's interface via active processes such as kinase-controlled phosphorylation. The latter has been observed experimentally during certain stages of cell division in {\em C. elegans} \cite{Folk21}.
Multiplying both sides of equation (\ref{radCL}) by $R_i^2$ and summing over $i$ yields
\begin{equation}
\frac{dV_{\rm drop}}{dt}=4\pi \kappa\left (\frac{\sum_{i=1}^N R_i^2}{R_c}-\sum_{i=1}^NR_i\right ).
\end{equation}
Conservation of $V_{\rm drop}$ now implies that
\begin{equation}
\label{Rc3D}
R_c(t)=\frac{\sum_{i=1}^N R_i^2(t)}{\sum_{i=1}^NR_i(t)}
\end{equation}
and, hence,
\begin{equation}
\label{phin2}
\Delta(t) =\frac{\ell_c\phi_a\sum_{i=1}^NR_i(t)}{\sum_{i=1}^NR_i^2(t)} .
\end{equation}
Eq, (\ref{phin2}) differs significantly from the condition (\ref{phin}) of classical Ostwald ripening in 3D.

In the following sections we derive the above mean-field model and obtain higher-order corrections that take into account rapid spatial variations in the concentration around each condensate. We proceed by replacing the uniform concentration $\overline{\phi}$ in equation (\ref{JRCL}) with the local concentration $\phi(R)$. Equation (\ref{JRCL}) now becomes \begin{equation}
\label{RB0}
D\phi'(R)=\kappa_0\left (\phi(R) -\phi_a-\frac{\phi_a \ell_c}{R}\right ).
\end{equation}
Note that, mathematically speaking, if $\phi_a=0$ then equation (\ref{RB0}) would reduce to a classical Robin or radiation boundary condition. Each droplet interface would act as a partially absorbing boundary and the resulting system would be unphysical, since the droplets would simply grow until only the dense phase exists. As we show in subsequent sections, the mixed boundary condition (\ref{RB0}) supports a non-trivial steady-state solution for $\phi$, which can then be used to determine the kinetics of droplet growth and shrinkage on a longer time scale.

\setcounter{equation}{0} 
\section{Asymptotic analysis in 3D}

Consider a set of $N$ small partially absorbing spherical condensates $\calU_k\subset \Omega \subset\R^3$, $k=1,\ldots,N$, as illustrated in Fig. \ref{fig1}(a). We fix the length and time scales by setting $D=1$ and $L\equiv |\Omega|^{1/3}=1$, and assume that $R_i=\epsilon \rho_i$ with $\rho_i=O(1)$ and $0<\epsilon \ll 1$. (Another important small parameter is the ratio of the dilute and dense concentrations $\phi_a/\phi_b$.) We also take the condensates to be well-separated from each other and the boundary $\partial \Omega$. 
The concentration $\phi$ of the dilute phase outside the droplets satisfies the quasi-static diffusion equation
\begin{subequations}
\label{diff}
\begin{equation}
\nabla^2 \phi=0,\quad \x \in \Omega\backslash \cup_{j=1}^N\calU_{j},
\end{equation}
supplemented by the boundary conditions
\begin{align}
\label{diff2}
 \nabla \phi \cdot\n &=0 \,\quad \x\in  \partial \Omega, \\
D\nabla \phi(\x)\cdot \n_j&=\kappa_0\left (\phi(\x) -\phi_a-\frac{\phi_a \ell_c}{|\x-\x_j|}\right ),\quad \x \in \partial \calU_j.
\end{align}
\end{subequations}
The vector $\n_j$ denotes the unit normal to the surface $\partial \calU_j$ that is directed outward from the interior of $\calU_j$.

First, consider the inner solution around the $j$th droplet,
\begin{equation}
\Phi_j(\y)=\phi(\x_j+\epsilon \y),\quad \y=\epsilon^{-1}(\x-\x_j),
\end{equation}
where we have introduced stretched coordinates and replaced the domain $\Omega $ by $\R^3$, see Fig. \ref{fig1}(b). 
The inner solution satisfies the diffusion equation
\begin{subequations}
\begin{equation}
\nabla_{\y}^2 \Phi_j=0 \mbox{ for all } |\y|>\rho_j, \end{equation}
with
\begin{align}
D\nabla_{\y} \Phi_j(\y)\cdot \n_j&=\epsilon \kappa_0\left (\Phi_j(\y) -\phi_a-\frac{\phi_a \ell_c}{\epsilon |\y|}\right ),\quad |\y|=\rho_j.
\end{align}
\end{subequations}
Following Ref. \cite{Bressloff20a}, we perform the rescaling $\ell_c=\epsilon \ell_c'$. (This is also consistent with the experimental observation that the intrinsically disordered protein MEG-3 forms assemblies on the surface of P granules, resulting in a lowering of the surface tension \cite{Folk21}. This is analogous to the formation of inorganic Pickering emulsions \cite{Ramsden,Pickering}.) Introducing spherical polar coordinates with $\rho=|\y|$, we have
\begin{subequations}
\label{symin}
\begin{align}
\frac{1}{\rho^2}\frac{d}{d\rho}\rho^2\frac{d\Phi_j}{d\rho}&=0,\quad \rho_j < \rho <\infty,\\
 D\Phi_j'(\rho_j)&=\epsilon \kappa_0\left (\Phi_j(\rho_j) -\phi_a-\frac{\phi_a \ell_c'}{\rho_j}\right )  .
 \end{align}
\end{subequations}

Next, the far-field behavior of the inner solution around each droplet has to be matched with the near-field behavior of the outer solution, see Fig. \ref{fig1}(c). This is achieved by expanding both the inner and outer solutions as power series in $\epsilon$ along analogous lines to Refs. \cite{Cheviakov11,Bressloff20a}. It turns out that the outer solution has an expansion of the form
\begin{equation}
\phi(\x)=\phi_{\infty} +\epsilon^2 \phi_1(\x)+\ldots,
\end{equation}
with
 \begin{equation}
 \label{3Dphi1}
\nabla^2 \phi_1=0,\quad \x\in \Omega\backslash \{\x_1,\ldots,\x_N\}, \ \nabla\phi_1\cdot \n=0,\ \x \in \partial \Omega,
\end{equation}
and $\phi_1(\x)$ singular as $\x\rightarrow \x_j$ for $j=1,\ldots,N$. Introducing a corresponding $\epsilon$-expansion of the inner solution around the $j$th droplet
\begin{equation}
\label{ephi}
\Phi_j(\rho)=\phi_{\infty}+\epsilon \Phi_{j,1}(\rho)+\epsilon^2 \Phi_{j,2}(\rho)+\ldots
\end{equation}
and substituting into equations (\ref{symin}) gives 
\begin{subequations}
\label{3Dinner0}
\begin{align}
\frac{1}{\rho^2}\frac{d}{d\rho}\left [\rho^2\frac{d\Phi_{j,1}}{d\rho}\right ]&=0,\quad \rho_j < \rho <\infty,\\
 D\Phi_{j,1}'(\rho_j)&=\kappa_0\left (\Delta-\frac{\phi_a \ell_c'}{\rho_j}\right )  ,\quad
\Phi_{j,1} \rightarrow 0\mbox{ as } \rho\rightarrow \infty,
 \end{align}
\end{subequations}
and
\begin{subequations}
\label{3Dinnerp1}
\begin{align}
\frac{1}{\rho^2}\frac{d}{d\rho}\left [\rho^2\frac{d\Phi_{j,2}}{d\rho}\right ]&=0,\quad \rho_j < \rho<\infty,\\
 D\Phi_{j,2}'(\rho_j)&=\kappa_0 \Phi_{j,1}(\rho_j)  ,\quad
\Phi_{j,2} \rightarrow \phi_{j,1}^{\rm reg}\mbox{ as } \rho\rightarrow \infty .
 \end{align}
\end{subequations}
Here ${\phi}_{j,1}^{\rm reg} $ denotes the non-singular part of $\phi_1(\x)$ as $\x\rightarrow \x_j$. Finally, equation (\ref{3Dphi1}) is supplemented by the matching condition 
$\epsilon \phi_1\sim \Phi_{j,1}$ as $\x\rightarrow \x_j$.

We now proceed iteratively. First, equations (\ref{3Dinner0}) have the solution 
\begin{equation}
\label{Phi1}
\Phi_{j,1}(\rho)= -\frac{\kappa_0\rho_j}{D}\frac{\rho_j}{\rho}\left (\Delta -\frac{\phi_a\ell_c'}{\rho_j}\right ).
\end{equation}
The outer solution $\phi_1$ of equation (\ref{3Dphi1}) is determined by introducing the Neumann Green's function $G(\x,\x')$, which is uniquely defined by
\begin{subequations}
\label{G}
\begin{equation}
\nabla^2 G(\x,\x') =\frac{1}{|\Omega|}-\delta(\x-\x'),\quad \x,\x' \in \Omega,
\end{equation}
and
\begin{equation}
\nabla G\cdot \n =0 \mbox{ on } \partial \Omega, \quad \int_{\Omega}G(\x,\x')d\x=0
\end{equation}
\end{subequations}
for fixed $\x'$.
Note that $G$ has the decomposition
\begin{equation}
G(\x,\x')=\frac{1 }{4\pi|\x-\x'|}+H(\x,\x'),
\end{equation}
where $H$ is the nonsingular part of the Green's function. 
The solution of equation (\ref{3Dphi1}) can then be written as
\begin{equation}
\label{outer1}
\phi_1(\x)\approx -4\pi  \sum_{j=1}^N\frac{\kappa_0\rho_j^2}{D}\left (\Delta -\frac{\phi_a\ell_c'}{\rho_j}\right )G(\x,\x_j)
\end{equation}
for $\x \notin \{\x_j,\, j=1,\ldots,N\}$ so that
\begin{align}
\nabla^2 \phi_1(\x) &\approx -4\pi  \sum_{j=1}^N\frac{\kappa_0\rho_j^2}{D}\left (\Delta -\frac{\phi_a\ell_c'}{\rho_j}\right )\nabla^2 G (\x,\x_j)=-\frac{4\pi}{|\Omega|}  \sum_{j=1}^N\frac{\kappa_0\rho_j^2}{D}\left (\Delta -\frac{\phi_a\ell_c'}{\rho_j}\right ).
\end{align}
Hence, the $O(\epsilon^2)$ term in the expansion of the outer solution satisfies the steady-state diffusion equation if and only if
$\sum_{j=1}^N\rho_j [\rho_j \Delta -{\phi_a}\ell_c']=0$, that is,
\begin{equation}
\phi_{\infty} =\phi_a\left (1+\frac{\ell_c' \sum_{j=1}^N \rho_j}{\sum_{j=1}^N\rho_j^2} \right ).
\end{equation}
This recovers equation (\ref{phin2}).
Finally, we determine the $O(\epsilon^2)$ correction to the inner solution by substituting the regular part of $\phi_1(\x) $ as $\x\rightarrow \x_j$ into equation (\ref{3Dinnerp1}):
\begin{equation}
\Phi_{j,2}(\rho)=\phi_{1,j}^{\rm reg }+\left (\frac{\kappa_0\rho_j}{D}\right)^2 \frac{\rho_j}{\rho}\left (\Delta -\frac{\phi_a\ell_c'}{\rho_j}\right ),
\end{equation}
with
\begin{align}
{\phi}_{1,j}^{\rm reg }&=-4\pi\sum_{k\neq j}\frac{\kappa_0\rho_k^2}{D}\left (\Delta -\frac{\phi_a\ell_c'}{\rho_k}\right )G (\x_j,\x_k)-4\pi \frac{\kappa_0\rho_j^2}{D}\left (\Delta -\frac{\phi_a\ell_c'}{\rho_j}\right )H(\x_j,\x_j).
\label{Lam}
\end{align}

Combining our various results, we obtain the following asymptotic expansion of the inner solution around the $j$th droplet:
\begin{align}
\Phi_j(\rho)&\sim \phi_{\infty} +\epsilon^2 {\phi}_{1,j}^{\rm reg }  -\epsilon \frac{\kappa_0\rho_j}{D}\frac{\rho_j}{\rho}\left (1-\epsilon\frac{\kappa_0\rho_j}{D}\right )\left (\Delta -\frac{\phi_a\ell_c'}{\rho_j}\right )+O(\epsilon^3).  \end{align}
Given the quasi-static inner solution, we can now write down the dynamical evolution of the radius $R_j(t)$ on longer time scales, which we express in terms of the rescaled variable $\rho_j$:
\begin{align}
\frac{dR_j}{dt}&=\frac{D}{\phi_b}\left .\frac{d\Phi_j(\rho)}{\epsilon d\rho}\right |_{\rho=\rho_j} \sim  \frac{\kappa_0}{\phi_b}\left (\Delta -\frac{\phi_a\ell_c'}{\rho_j}\right )\left (1- \epsilon \frac{\kappa_0\rho_j}{D}\right ).
\label{3Ddyn}
\end{align}
A number of observations follow from equation (\ref{3Ddyn}).
\medskip

\noindent (i) Dropping the factor $(1-\epsilon \kappa_0\rho_j/D)$ on the right-hand side of equation (\ref{3Ddyn}) and using the unscaled variables $R_j,\ell_c$, we recover equation (\ref{radCL}).
The asymptotic analysis generates higher-order corrections to this mean-field result. Note, in particular, that global aspects of the system as encoded in the Green's function enter at $O(\epsilon^2)$ on the right-hand side of equation (\ref{3Ddyn}). That is, $D\Phi_{j,3}'(\rho_j)=\kappa_0 \Phi_{j,2}(\rho_j)$ and $\Phi_{j,2}(\rho_j)$ depends on $\phi_{1,j}^{\rm reg }$, see equation (\ref{Lam}).
\medskip

\noindent (ii) Equation (\ref{3Ddyn}) implies that the rate of growth and shrinkage of a droplet depends on the two parameters $\kappa_0$ and $\phi_a/\phi_b$. Since $\phi_a/\phi_b\ll 1$, the quasi-static diffusion approximation will be valid provided that $\kappa_0$ is sufficiently small.
\medskip

\noindent (iii) The asymptotic analysis of 3D Ostwald ripening leads to a kinetic equation of the form \cite{Bressloff20a}
\begin{align}
\epsilon\frac{dR_j}{dt}
&\sim \frac{D}{\rho_j\phi_b} \left (\Delta -\frac{\phi_a\ell_c'}{\rho_j}+\epsilon \phi_{1,j}^{\rm reg}\right ),
\label{3Dost}
\end{align}
with
\begin{align}
{\phi}_{1,j}^{\rm reg }&=-4\pi\sum_{k\neq j} \rho_k \left (\Delta -\frac{\phi_a\ell_c'}{\rho_k}\right )G (\x_j,\x_k) -4\pi \rho_j \left (\Delta -\frac{\phi_a}{\rho_j}\right )H(\x_j,\x_j).
\label{Lamor}
\end{align}
The left-hand side of equation (\ref{Lamor}) is scaled by a factor $\epsilon$, which is consistent with the proposal that conversion-limited phase separation slows coarsening relative to classical Ostwald ripening.

\setcounter{equation}{0}
\section{Asymptotic analysis in 2D}

As far as we are aware, there is not yet any experimental evidence for conversion-limited phase separation in 2D. On the other hand, phase separation also appears to play a role in the formation and maintenance of synapses \cite{Zeng16,Bai21,Hosokawa21}, which could be idealized as 2D droplets. Therefore, in this section we carry out an asymptotic analysis of 2D conversion-limited phase separation. As we mentioned in the introduction, a major difference between 3D and 2D is the singularity structure of the associated Green's function. One consequence of this is that in 2D there are two small parameters $\epsilon $ and $\delta = -\epsilon \ln \epsilon$. (Note that $\delta \rightarrow 0$ as $\epsilon \rightarrow 0$ but converges to zero more slowly.) We proceed by carrying out an asymptotic expansion in $\epsilon$, with the coefficients  of the expansion depending non-perturbatively on $\delta$.

 The 2D version of the inner equations (\ref{symin}) is
\begin{subequations}
\label{2Dsymin}
\begin{align}
\frac{1}{\rho}\frac{d}{d\rho}\rho\frac{d\Phi_j}{d\rho}&=0,\quad \rho_j < \rho <\infty,\\
 D\Phi_j'(\rho_j)&= \epsilon \kappa_0\left (\Phi_j(\rho_j) -\phi_a-\frac{\phi_a \ell_c'}{\rho_j}\right )  .
 \end{align}
\end{subequations}
We expand the inner solution as the perturbation series 
\begin{equation}
\label{2Dephi}
\Phi_j(\rho)=\Phi_{j,0}(\rho;\delta)+\epsilon \Phi_{j,1}(\rho;\delta)+\ldots
\end{equation}
 Substituting into equation (\ref{2Dsymin}) implies that $\Phi_{j,0}(\rho;\delta)=A_{j,0}(\delta)$ and
\begin{align}
\Phi_{j,1}(\rho;\delta)&=A_{j,1}(\delta)  +\frac{\kappa_0 \rho_j}{D}\left (A_{j,0}(\delta) -\phi_a -\frac{\phi_a\ell_c'}{\rho_j}\right )\ln (\rho/\rho_j).\nonumber\end{align}
The outer solution is obtained by treating each droplet as a point source/sink, see Fig. \ref{fig1}(c). The resulting time-independent diffusion equation takes the form
\begin{subequations}
\begin{equation}
\nabla^2 \phi=0,\quad \x\in \Omega\backslash \{\x_1,\ldots,\x_N\}, \quad \nabla \phi\cdot \n =0,\ \x \in \partial \Omega,
\end{equation}
together with the matching condition
\begin{align}
\phi &\sim A_{j,0}(\delta)+\epsilon A_{j,1}(\delta)+ F(\rho_j;\delta) \delta +\epsilon F(\rho_j;\delta)\left [\ln (|\x-\x_j|)-\ln  \rho_j\right ]
\label{match}
\end{align}
\end{subequations}
as $\x\rightarrow \x_j$, with
\begin{equation}
F(\rho_j;\delta)=\frac{\kappa_0 \rho_j}{D}\left (A_{j,0}(\delta)-\phi_a -\frac{\phi_a\ell_c'}{\rho_j}\right ).
\label{F}
\end{equation}
 Let $G(\x,\x')$ denote the 2D version of the Neumann Green's function, which satisfies equations (\ref{G}) and has the decomposition
 \begin{equation}
G(\x,\x')=-\frac{\ln|\x-\x'|}{2\pi}+H(\x,\x'),
\end{equation}
where $H$ is now the nonsingular part of the 2D Green's function. Given $G(\x,\x')$, we introduce the ansatz 
\begin{equation}
\label{outer}
\phi(\x)\approx \phi_{\infty}-2\pi  \epsilon  \sum_{i=1}^NF(\rho_i;\delta)G(\x,\x_i)
\end{equation}
for $\x \notin \{\x_j,\, j=1,\ldots,N\}$
for some constant $\phi_{\infty}$. Observe that 
\begin{align}
\nabla^2 \phi(\x) &\approx -2\pi \epsilon \sum_{i=1}^NF(\rho_i;\delta)\nabla^2 G (\x,\x_i) = -\frac{2\pi \epsilon}{|\Omega|}  \sum_{i=1}^NF(\rho_i;\delta).
\end{align}
Hence, the outer solution satisfies the steady-state diffusion equation if and only if
\begin{equation}
\label{sumA}
\sum_{j=1}^NF(\rho_j;\delta)=0.
\end{equation}

The outer solution has the singular behavior 
\begin{align}
\phi(\x)&\rightarrow \phi_{\infty} +\epsilon  F(\rho_j;\delta)\ln|\x-\x_j|-2\pi \epsilon F(\rho_j;\delta) H(\x_j,\x_j)\nonumber \\
&\qquad -2\pi \epsilon  \sum_{i\neq j}^NF(\rho_i;\delta)G(\x_j,\x_i)
\end{align}
as $\x\rightarrow \x_j$.
Comparison with the asymptotic limit in equation (\ref{match}) yields the self-consistency conditions
\begin{align}
\label{Ai}
& 2\pi \epsilon H(\x_j,\x_j)F(\rho_j;\delta)+ 2\pi \epsilon \sum_{i\neq j} F(\rho_i;\delta)G(\x_j,\x_i)\nonumber
 \\
&\hspace{2cm} =\phi_{\infty}-A_{j,0}(\delta)-\epsilon A_{j,1}(\delta) +\epsilon F(\rho_j;\delta)\ln \rho_j-F(\rho_j;\delta)\delta \end{align}
for $ j=1,\ldots,N$. Hence, equating terms of $O(1)$ gives
\begin{equation}
A_{j,0}(\delta)=\phi_{\infty}-\frac{\kappa_0 \rho_j}{D}\left (A_{j,0}(\delta)-\phi_a -\frac{\phi_a\ell_c'}{\rho_j}\right )\delta ,
\end{equation}
which can be rearranged so that
\begin{equation}
\label{Aj0}
A_{j,0}(\delta)=\frac{\phi_{\infty}+\left (\phi_a +\frac{\dis \phi_a\ell_c'}{\dis \rho_j}\right )\delta}{1+\frac{\dis \kappa_0 \rho_j \delta}{\dis D}}.
\end{equation}
Similarly, at $O(\epsilon)$ we have
\begin{align}
A_{j,1}(\delta)&=-2\pi \bigg[ F(\rho_j;\delta)H(\x_j,\x_j)+ \sum_{i\neq j} F(\rho_i;\delta)G(\x_j,\x_i)  -F(\rho_j;\delta)\ln \rho_j\bigg ]\nonumber \\
&\qquad \times \left [1+\frac{\dis \kappa_0 \rho_j \delta}{\dis D}\right ]^{-1}.
\end{align}
The final factor follows from going to $O(\epsilon^2)$.

We can now use the inner solution to determine the long-time evolution of the radius $R_j$:
\begin{align}
\frac{dR_j}{dt}&\sim  \frac{\kappa_0}{\phi_b} \left (A_{j,0}(\delta)-\phi_a -\frac{\phi_a\ell_c'}{\rho_j}\right )\nonumber \\
&\sim   \frac{\kappa_0}{\phi_b} \left (\Delta -\frac{\phi_a\ell_c'}{\rho_j}\right )  +  \frac{\kappa_0 \delta}{\phi_b} \left [ \frac{\phi_a +\frac{\dis \phi_a\ell_c'}{\dis \rho_j}-\frac{\dis \kappa_0\rho_j\phi_{\infty}}{\dis D}}{1+\frac{\dis \kappa_0 \rho_j \delta}{\dis D}}\right ].
\label{2Ddyn}
\end{align}
A number of observations follow from equation (\ref{2Ddyn}).
\medskip

\noindent (i) The first term on the second line of equation (\ref{2Ddyn}) is identical to the leading order contribution of the corresponding 3D result (\ref{3Ddyn}). However, in contrast to 3D, there are $\delta$-dependent corrections to the leading order dynamics given by the second term. In addition, these corrections depend non-perturbatively on $\delta$. 
\medskip

\noindent (ii) As in 3D, the rate of growth and shrinkage depends on the two parameters $\kappa_0$ and $\phi_a/\phi_b$.
\medskip

\noindent (iii) Multiplying both sides of the first line of equation (\ref{2Ddyn}) by $2\pi R_j$ and summing with respect to $j$ implies that the total area of the droplets, $A_{\rm drop}=\pi \sum_{j=1}^NR_j^2$ ,  satisfies 
\begin{equation}
\frac{dA_{\rm drop}}{dt}=\epsilon^2\sum_{j=1}^NF(\rho_j;\delta),
\end{equation}
with $F(\rho_j;\delta)$ defined by equation (\ref{F}). It follows that the self-consistency condition (\ref{sumA}) is equivalent to conservation of total droplet area. Moreover, under the leading order approximation $A_{j,0}(\delta)\approx \phi_{\infty}$, equation (\ref{sumA}) simplifies to the condition (in orginal variables)
$R_c(t)=\sum_{j=1}^NR_j(t)$.
Hence, at the given level of approximation, the supersaturation for 2D conversion-limited phase separation is related to the droplet radii in an identical fashion to classical Ostwald ripening in 3D.
\medskip

\noindent (iv) The asymptotic analysis of 2D Ostwald ripening leads to a kinetic equation of the form \cite{Kavanagh14,Bressloff20a}
\begin{align}
\epsilon\frac{dR_j}{dt}
&\sim \frac{D}{\rho_j\phi_b} \nu A_j(\nu),\quad \nu =-\frac{1}{\ln \epsilon},
\label{2Dost}
\end{align}
with the coefficients $A_j(\nu)$ given by
\begin{equation}
\label{2DAi2}
A_i(\nu)=\sum_{j=1}^N[{\bf I}+\nu {\bf M}]^{-1}_{ij}\left (\Delta -\frac{\phi_a\ell_c'}{\rho_j}\right ),
\end{equation}
and ${\bf M}$ an $N\times N$ matrix with elements
\begin{align}
M_{jj}&=2\pi R (\x_j,\x_j)-\ln \rho_j,\quad M_{ji}=2\pi G (\x_j,\x_i),\, j\neq i.
\label{M}
\end{align}
Equation (\ref{2DAi2}) is clearly non-perturbative with respect to $\nu$. Finally, in the case of 2D Ostwald ripening, the supersaturation is determined by the harmonic mean of the radii \cite{Kavanagh14,Bressloff20a}:
\begin{equation}
\label{2Dphin2}
\Delta(t) =\frac{\ell_c\phi_a}{R_{\rm harm}(t)} ,\quad R_{\rm harm}(t)=\frac{1}{N}\sum_{j=1}^N\frac{1}{R_j(t)}.
\end{equation}

\section{Conversion-limited phase separation in a concentration gradient}

\begin{figure}[b!]
  \centering
  \includegraphics[width=13cm]{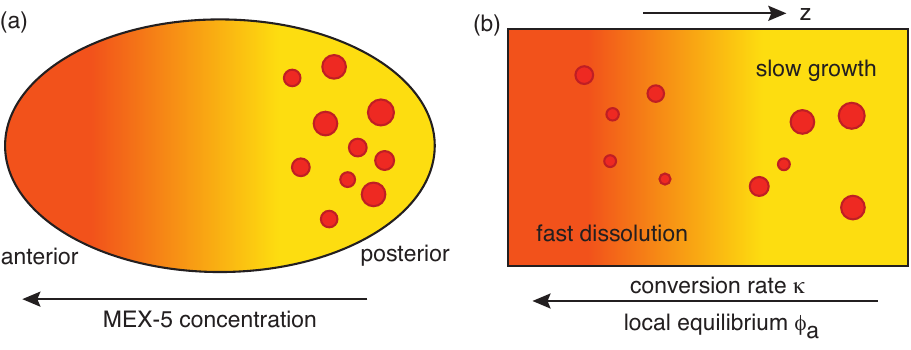}
  \caption{ (a) Schematic diagram illustrating the distribution of P granules following zygote polarization driven by an MEX-5 protein concentration gradient. (b) Schematic diagram of fast anterior dissolution and slow posterior growth in an idealized cylindrical domain with gradients in $\kappa$ and $\phi_a$.}
  \label{fig3}
\end{figure}

P granules play an important role in the life cycle of {\em C. elegans} \cite{Seydoux18}. For most of the life cycle, P granules are found in a cytoplasmic neighborhood of the nucleus. On the other hand, during cell fertilization (the oocyte-to-zygote transition), P granules redistribute throughout the cytoplasm following a rapid cycle of dissolution and condensation. A second dynamical phase occurs during zygote polarization, where P granules dissolve in the anterior cytoplasm while undergoing enhanced condensation in the posterior cytoplasm, see Fig. \ref{fig3}(a). Experimental studies suggest that this polarization is driven by a concentration gradient in the protein MEX-5 along the anterior-posterior axis. The latter is thought to generate corresponding spatial gradients in the conversion rate $\kappa_0$ and the local equilibrium $\phi_a$ (due to modifications in the interaction energy between solute and solvent).
The effects of the latter have been explored theoretically within the context of Ostwald ripening \cite{Weber17,Bressloff20b}. In addition, numerical simulations of P granule polarization have been carried out in Ref. \cite{Folk21}, using a spatially regulated model of conversion-limited phase separation.

Since intracellular protein concentration gradients are found in a wide range of cells undergoing polarization, it is of interest to explore different theoretical scenarios for the segregation of biological concentrates. In this section we use the asymptotic theory developed in section 3 to investigate the effects of spatial gradients in $\kappa_0$ and $\phi_a$ on late-stage 3D conversion-limited phase separation. For simplicity, we assume that the total volume of the dense phase is conserved and that there is still a separation of time scales between diffusion and droplet growth/shrinkage. (In the particular example of P granule  dynamics during zygote polarization, the total droplet volume tends to increase \cite{Folk21}. Moreover, dissolution and assembly occurs over several minutes so that the separation of time scales may be weaker than in other stages of the life cycle.) Let $\Omega\in \R^3$ be a cylinder whose axis is along the $z$ direction with $z\in [0,L]$. Suppose that there is a spatially varying conversion rate $\kappa(z)$ and a spatially varying local equilibrium $\phi_a(z)$ that take the form of linear gradients in the $z$-direction, see Fig. \ref{fig3}(b):
\begin{equation}
\kappa(z)=\kappa_0  -\beta z,\quad \phi_a(z)= \phi_{a,0} -\beta' z,\quad \quad \beta,\beta' >0.
\end{equation}
The asymptotic analysis of section 3 can be extended to take into account these heterogeneities. The main difference is that the inner solution around each droplet is no longer spherically symmetric. Introduce stretched coordinates around the $j$-th droplet centered at $\x_j=(x_j,y_j,z_j)$ according to
\begin{equation}
x=x_j+\epsilon \rho\sin \theta \cos \phi,\ y=y_j+\epsilon \rho\sin \theta \sin \phi, \ z=z_j+\epsilon \rho\cos \theta ,\quad \rho\geq \rho_j.
\end{equation}
This leads to a modified version of equations (\ref{symin}) for $\Phi_j=\Phi_j(\rho,\theta)$: 
\begin{subequations}
\label{inner3D}
\begin{align}
&\frac{1}{\rho^2}\frac{\partial}{\partial \rho}\rho^2\frac{\partial \Phi_j}{\partial \rho}+\frac{1}{\rho^2\sin \theta}\frac{\partial}{\partial \theta}\left [\sin \theta \frac{\partial \Phi_j}{\partial \theta }\right ]=0,\quad \rho_j < \rho <\infty,\\
& D\frac{\partial \Phi_j}{\partial \rho}=\epsilon (\kappa_j-\epsilon\beta  \rho_j\cos \theta)\left [\Phi_j -(\phi_{j}-\epsilon \beta'  \rho_j\cos \theta)\left (1+\frac{\ell_c'}{\rho_j}\right )\right ]  ,\quad \rho=\rho_j
 \end{align}
\end{subequations}
 for $\kappa_j=\kappa_0 -\beta z_j$ and $\phi_j=\phi_{a,0}-\beta' z_j$.
 
Since, the $\theta$-dependent terms are $O(\epsilon^2)$, we introduce the $\epsilon$-expansion
\begin{equation}
\label{ephi2}
\Phi_j(\rho,\theta)=\phi_{\infty}+\epsilon \Phi_{j,1}(\rho)+\epsilon^2 \Phi_{j,2}(\rho,\theta)+\ldots
\end{equation}
The $O(\epsilon)$ inner equation is
\begin{subequations}
\begin{align}
\frac{1}{\rho^2}\frac{d}{d\rho}\left [\rho^2\frac{d\Phi_{j,1}}{d\rho}\right ]&=0,\quad \rho_j < \rho <\infty,\\
 D\Phi_{j,1}'(\rho_j)&=\kappa_j\left (\Delta_j-\frac{\phi_j \ell_c'}{\rho_j}\right )  ,\quad
\Phi_{j,1} \rightarrow 0\mbox{ as } \rho\rightarrow \infty,
 \end{align}
 \end{subequations}
 where $\Delta_j=\phi_{\infty}-\phi_j$.
It follows that $\Phi_{j,1}$ is given by equation (\ref{Phi1}) with $\kappa_0\rightarrow \kappa_j$ and $\phi_a\rightarrow \phi_j$. Hence, the $O(\epsilon^2)$ contribution to the outer solution becomes
 \begin{equation}
\label{newouter1}
\phi_1(\x)\approx -4\pi  \sum_{j=1}^N\frac{\kappa_j\rho_j^2}{D}\left (\Delta_j -\frac{\phi_j\ell_c'}{\rho_j}\right )G(\x,\x_j).
\end{equation}
 The non-trivial modification of the inner solution occurs at $O(\epsilon^2)$ with
 \begin{subequations}
 \label{inner3Da}
\begin{align}
&\frac{1}{\rho^2}\frac{\partial}{\partial \rho}\rho^2\frac{\partial \Phi_{j,2}}{\partial \rho}+\frac{1}{\rho^2\sin \theta}\frac{\partial}{\partial \theta}\left [\sin \theta \frac{\partial \Phi_{j,2}}{\partial \theta }\right ]=0,\quad \rho_j < \rho<\infty,\\
& \left .D\frac{\partial \Phi_{j,2}}{\partial \rho}\right |_{\rho=\rho_j}= \kappa_j\Phi_{j,1}(\rho_j)-\beta  \rho_j\cos \theta\left (\Delta_j-\frac{\phi_j \ell_c'}{\rho_j}\right ) +\beta'   \kappa_j\rho_j\cos \theta\left (1+\frac{\ell_c'}{\rho_j}\right ),  \\
&\Phi_{j,2}\rightarrow \phi_{j,1}^{\rm reg}   \mbox{ as } \rho \rightarrow \infty. 
 \end{align}
\end{subequations}
The solution to equation (\ref{inner3Da}a) takes the from
\begin{equation}
\label{Leg}
\Phi_{j,2}(\rho,\theta)=\sum_{n=0}^{\infty}\left (A_n\rho^n+B_n\rho^{-n-1}\right )P_n(\cos \theta),
\end{equation}
where $P_n(\cos \theta)$ are the Legendre polynomials. The asymptotic condition  (\ref{inner3Da}c) implies that $A_n=0$ for all $n>0$ and $A_0=\phi_{j,1}^{\rm reg}$. Substituting into equation (\ref{inner3Da}b) yields
\begin{align}
 D\sum_{n=0}^{\infty}(n+1)B_n\rho_j^{-n-2}P_n(\cos \theta)&=\kappa_j\Phi_{j,1}(\rho_j)+\rho_j\beta \cos \theta \left (\Delta_j -\frac{\phi_j\ell_c'}{\rho_j}\right )\nonumber \\
&\quad -  \kappa_j\beta' \rho_j\cos \theta\left (1+\frac{\ell_c'}{\rho_j}\right ) .
\end{align}
From the orthogonality relation for Legendre polynomials we obtain the result
\begin{align}
B_0&=\left (\frac{\kappa_j \rho_j}{D}\right )^2\rho_j\left (\Delta_j -\frac{\phi_j\ell_c'}{\rho_j}\right ),\quad B_1 =\frac{ \rho_j^4\beta}{2D} \left (\Delta_j -\frac{\phi_j\ell_c'}{\rho_j}\right )-\frac{\kappa_j\rho_j^4\beta'}{2D} \left (1+\frac{\ell_c'}{\rho_j}\right ).
\end{align}
We conclude that the inner solution around the $j$-th droplet is to $O(\epsilon^2)$
\begin{align}
\label{Nin}
\Phi_j(\rho)&\sim a_j(\epsilon)  -\epsilon b_j(\epsilon)\frac{\rho_j}{\rho}+\epsilon^2c_j \left (\frac{\rho_j}{\rho}\right )^2\cos \theta,  \end{align}
where
\begin{align}
a_j(\epsilon)&= \phi_{\infty} +\epsilon^2 {\phi}_{1,j}^{\rm reg } ,\quad b_j(\epsilon)=\frac{\kappa_j\rho_j}{D}\left (1-\epsilon\frac{\kappa_j\rho_j}{D}\right )\left (\Delta_j -\frac{\phi_j\ell_c'}{\rho_j}\right ),\nonumber \\
c_j&=\frac{ \rho_j^2\beta}{2D} \left (\Delta_j -\frac{\phi_j\ell_c'}{\rho_j}\right )-\frac{\kappa_j\rho_j^2\beta'}{2D} \left (1+\frac{\ell_c'}{\rho_j}\right ).
\end{align}

On longer time scales, the droplets can grow, drift or deform due to the normal fluxes of solute at the interface modifying the location of the interface.
The local displacement $\delta R_j(\theta,t)$ of the $j$-th droplet interface is determined by the normal flux into the surface element $d\sigma_j=R_j d\Gamma$ where $d\Gamma$ denotes an infinitesimal solid angle. Conservation of solute implies that
\begin{equation}
\frac{\partial R_j(\theta,\phi,t)}{\partial t}\phi_b d\sigma_j=D\left. \frac{\partial \Phi_i( \rho,\theta)}{\epsilon\partial \rho}\right |_{\rho=\rho_j}d\sigma_j .
\end{equation}
From equation (\ref{Nin}), we have
\begin{eqnarray}
\left . \frac{\partial \Phi_i( \rho,\theta)}{\epsilon\partial \rho}\right |_{\rho=\rho_j}=  \frac{b_j(\epsilon)}{\rho_j}-\frac{2\epsilon c_j}{\rho_j} \cos \theta .
 \label{J}
\end{eqnarray}
Assuming that the droplet maintains an approximately spherical shape, we can define the change in radius according to
\begin{equation}
\delta R_j(t)=\frac{1}{\pi}   \int_0^{\pi} \delta R_j(\theta,t)d\theta.
\end{equation}
That is,
\begin{eqnarray}
\label{rho}
\frac{dR_j}{dt}\sim\frac{D}{\pi \phi_b }\int_0^{\pi}\left [\frac{b_j(\epsilon)}{\rho_j}-\frac{2\epsilon c_j}{\rho_j} \cos \theta\right ]d\theta=\frac{D}{\phi_b } \frac{b_j(\epsilon)}{\rho_j}\sim  \frac{\kappa_j}{\phi_b}\left (\Delta_j -\frac{\phi_j\ell_c'}{\rho_j}\right )+O(\epsilon),
\end{eqnarray}
and we recover equation (\ref{3Ddyn}) with $\kappa_0\rightarrow \kappa_j=\kappa_0-\beta z_j$ and $\phi_a\rightarrow \phi_j=\phi_{a,0}-\beta'z_j$. Since $\beta,\beta'>0$, it follows that the conversion factor $\kappa_j$ and the critical radius $\phi_j\ell_c /\Delta_j$ for droplet growth are both decreasing functions of droplet position $z_j$. Hence, droplets on the left-hand side of the cylindrical domain in Fig. \ref{fig3}(b) will tend to shrink, whereas droplets on the right-hand side will tend to grow. Moreover, dissolution will occur more quickly than assembly. There is also a second-order effect of the spatial gradients, namely, that droplets tend to slowly drift  in the $z$ direction according to
\begin{align}
\frac{d z_j}{dt}&=\frac{1}{\pi}\int_0^{\pi} \frac{\partial R_j(\theta,t)}{\partial t}\cos \theta d\theta=\frac{D}{\pi \phi_b }\int_0^{\pi} \left [\frac{b_j(\epsilon)}{\rho_j}-\frac{2\epsilon c_j}{\rho_j} \cos \theta\right ]\cos \theta d\theta\nonumber\\
&=-\frac{\epsilon D}{\phi_b }\frac{c_j}{\rho_j} =-\frac{\epsilon \rho_j\beta}{2\phi_b} \left (\Delta_j -\frac{\phi_j\ell_c'}{\rho_j}\right )+\frac{\epsilon \kappa_j\rho_j\beta'}{2\phi_b} \left (1+\frac{\ell_c'}{\rho_j}\right ).
\label{speed}
\end{align}
Equation (\ref{speed}) implies that droplets tend to move towards decreasing values of the local equilibrium $\phi_a$, consistent with previous studies of droplet ripening \cite{Weber17,Bressloff20b}. Interestingly,  the dependence of the drift velocity on the gradient of $\kappa$ depends on whether or not the radius is larger or smaller than the local critical radius. That is, locally growing (shrinking) droplets moves towards increasing (decreasing) values of the conversion rate $\kappa$.

\section{Discussion} 

In this paper, we used the mathematics of diffusion in singularly perturbed domains to analyze a model of conversion-limited phase separation. The latter is based on recent experimental and modeling studies of the slow coarsening of P granules in germ cells of  {\em C. elegans} \cite{Folk21,Lee21}. In contrast to classical diffusion-limited Ostwald ripening, the rate limiting step for the growth and shrinkage of high density droplets is the adsorption/desorption of solute at the droplet interfaces. Exploiting the fact that P granules are much smaller than the size of a cell, we carried out a matched asymptotic expansion of the solution to the quasi-static diffusion equation for the dilute volume fraction. This allowed us to derive the mean-field equations for droplet growth and shrinkage, together with higher-order corrections that depend on the geometry of the domain. These higher-order corrections also generated a slow drift of droplets in the presence of a spatial gradient.

One of the advantages of asymptotic analysis is that it can also be applied to 2D liquid-liquid phase separation, where mean field theory breaks down due to logarithmic singularities. In this paper we showed that in the case of conversion-limited phase separation, the asymptotic analysis involved a perturbation expansion in powers of $\epsilon$, with coefficients depending non-perturbatively on $\delta=-\epsilon \ln \epsilon$. (Recall that $\epsilon$ specifies the size of droplets relative to the size of the bulk domain $\Omega$.) This differs significantly from the analysis of 2D Ostwald ripening, which involves a summation over logarithmic terms involving powers of $\nu=-1/\ln \epsilon$ \cite{Kavanagh14,Bressloff20a}. One potential biological application of 2D coarsening is to the formation and maintenance of post-synaptic domains in the dendrites of neurons \cite{Zeng16,Bai21,Hosokawa21}. It would be interesting to determine whether or not there are analogs of conversion-limited phase separation and pickering proteins in such systems. 

Finally, another important issue for future consideration is deriving the interfacial boundary condition of conversion-limited phase separation from first principles. That is, equation (\ref{RB0}) is phenomenological in nature, rather than being based on the underlying biophysics. As highlighted in Refs. \cite{Folk21,Lee21}, the effective conversion rate is probably determined by a rugged energy landscape that arises from multi-valent interactions of various protein domains. This is also consistent with the observation that P granules act like high viscosity ``aging Maxwell fluids''  with glass-like properties \cite{Jawerth20}.

\end{document}